\documentclass[aps,prb,twocolumn,nobibnotes]{revtex4-1}  
\usepackage{dcolumn}
\usepackage{graphicx}
\usepackage{color}
\usepackage{amssymb}
\usepackage{amsmath}
\usepackage{bm}
\usepackage[colorlinks=true,linktocpage=true,breaklinks=true]{hyperref}

\begin{document}

\title{\bf Crossover from Diffusive to Superfluid Transport in Frustrated Magnets}

\author{V. M. L. D. P. Goli$^{1}$}
\author{Aur\'{e}lien Manchon$^{1,2}$}
\email{manchon@cinam.univ-mrs.fr}
\affiliation{$^1$Physical Science and Engineering Division (PSE), King Abdullah University of Science and Technology (KAUST), Thuwal 23955-6900, Saudi Arabia\\
$^2$Aix-Marseille Univ, CNRS, CINaM, Marseille, France}

\begin{abstract}
We investigate the spin transport across the magnetic phase diagram of a frustrated antiferromagnetic insulator and uncover a drastic modification of the transport regime from spin diffusion to spin superfluidity. Adopting a triangular lattice accounting for both nearest neighbor and next-nearest neighbor exchange interactions with easy-plane anisotropy, we perform atomistic spin simulations on a two-terminal configuration across the full magnetic phase diagram. We found that as long as the ground state magnetic moments remain in-plane, irrespective of whether the magnetic configuration is ferromagnetic, collinear or non-collinear antiferromagnetic, the system exhibits spin superfluid behavior with a device output that is independent on the value of the exchange interactions. When the magnetic frustration is large enough to compete with the easy-plane anisotropy and cant the magnetic moments out of the plane, the spin transport progressively evolves towards the diffusive regime. The robustness of spin superfluidity close to magnetic phase boundaries is investigated and we uncover the possibility for {\em proximate} spin superfluidity close to the ferromagnetic transition.
\end{abstract}

\maketitle

\section{Introduction}
\begin{figure}[t]
\includegraphics[width=9.0cm]{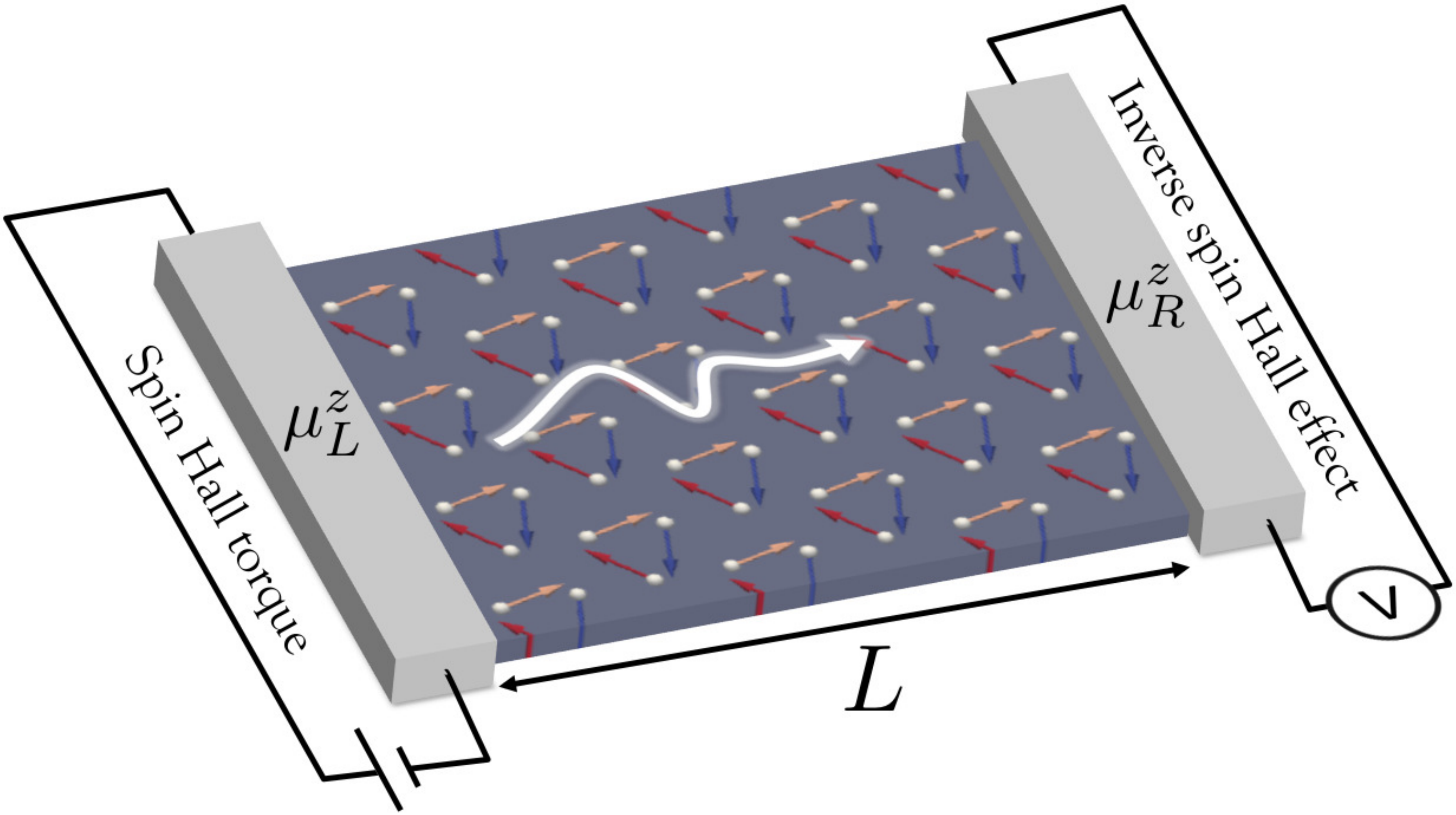}
\caption{The device setup: a non-collinear antiferromagnet of length $L$ is embedded between two heavy metal electrodes (grey). On the left electrode, a charge current is injected, which induces a spin accumulation $\mu_L^z$ at the interface with the antiferromagnet via spin Hall effect. This non-equilibrium spin accumulation induces a spin current in the antiferromagnet that is collected on the right electrode via inverse spin Hall effect. The ratio $\mu_R^z/\mu_L^z$ determines the efficiency of the device.\label{Fig1}}
\end{figure}

In the past few years, insulating magnets have been attracting increasing attention due to their ability for low-dissipation magnon transport, free from Joule heating \cite{Serga2010,Chumak2015}. Y$_3$Fe$_5$O$_{12}$ (YIG), an insulating collinear ferrimagnet, is the most widely used material for electrically induced magnonic transport as demonstrated experimentally in local \cite{Nakayama2013} and non-local device geometries \cite{Cornelissen2015}. In these systems however the magnon-mediated spin transport is in the diffusive regime, i.e., the transmitted signal decays exponentially with increasing the insulator thickness, $\sim e^{-\alpha qd}$, where $\alpha$ is the Gilbert damping, $q$ is the magnon wave vector and $d$ the magnetic layer thickness. In recent years, it has been proposed that in easy-plane ferromagnets spin transport enters a completely different regime, the spin superfluidity \cite{Sonin2010,Konig2001,Takei2014}. Spin superfluidity is the dissipationless transport of spin information in magnetically ordered systems, analogous to the conventional superfluidity realized in systems displaying broken U(1) rotational symmetry such as  $^4$He \cite{Kapitza1938,Allen1938,Volovik2003,Bunkov2013}. Similarly, ferromagnets with easy-plane anisotropy are characterized by a ground state with magnetic order lying in the plane, resulting in the breaking of U(1) symmetry \cite{Sonin2010, Konig2001,Kim2017f}. Hence spin superfluid transport is possible in these systems although long-range dipole interaction limits the phase coherence to less than a few hundred nanometers \cite{Skarsvag2015}. In a non-local device geometry consisting in an insulating magnet embedded between two metallic contacts, as depicted in Fig. \ref{Fig1}, the signal decays {\em algebraically} as a function of the magnetic insulator thickness \cite{Takei2014}, $\sim 1/(1+d/d_\alpha)$, where $d_\alpha=(\hbar/2\pi)(g^{\uparrow\downarrow}/\alpha S)$ with $g^{\uparrow\downarrow}$ being the spin mixing conductance characterizing the interfacial spin transmission/collection \cite{Brataas2006} and $S$ being the spin magnitude.\par

Antiferromagnetic insulators constitute appealing candidates to realized long-range spin transport due to their vanishing demagnetizing field and ultrafast dynamics \cite{Baltz2018}. In contrast to ferromagnetic insulators, the degenerate magnon transport yields no net spin transport in antiferromagnets (see discussion in Ref. \onlinecite{Ohnuma2013}). Nonetheless, this degeneracy can be lifted by injecting spin currents via, e.g., spin Hall effect\cite{Dyakonov1971b,Sinova2015} in a proximate heavy metal. The electron-driven spin current is then converted into a magnon-driven spin current \cite{Ohnuma2013,Khymyn2016} which, due to the bosonic nature of the magnons, can trigger Bose-Einstein condensation \cite{Giamarchi2008,Fjaerbu2017}. So far, only a few experimental works tackle directly the transport of spin information in antiferromagnets. Most previous works focus on antiferromagnet/ferromagnet bilayers, where the ferromagnet is used as a spin-injector. In those systems, the spin diffusion length of the antiferromagnet was reported to be only of a few nanometers in both insulating \cite{Wang2014d,Hahn2014,Qiu2018} and metallic \cite{Saglam2016,Frangou2016,Wen2019} systems. However in such systems, interfacial magnetic correlations play a crucial role that can hamper the magnon transmittivity \cite{Baldrati2018a,Jungfleisch2018}. Non-local injection of spin current following the scheme of Ref. \onlinecite{Cornelissen2015} has allowed for the demonstration that magnons propagate in collinear insulating antiferromagnets over micrometer length scales similarly to the best ferromagnetic systems \cite{Lebrun2018,Yuan2018}. \par

Theoretical studies on antiferromagnetic insulators have shown how spin superfluid transport can be realized \cite{Takei2014b}. For example, in the biaxial antiferromagnetic insulator NiO, non-local spin transport can be obtained up to a few micrometers \cite{Qaiumzadeh2017}. In this system, spin superfluid transport occurs at spin-flop transition by applying a magnetic field along the (weak) uniaxial anisotropy direction. Non-local spin transport has been experimentally observed in Cr$_2$O$_3$ by overcoming the lack of easy-plane anisotropy by applying the external magnetic field above the spin-flop transition, yielding spin transmission with algebraic decay up to 20 $\mu$m, thereby demonstrating spin superfluidity at 2 K \cite{Yuan2018}. Spin superfluidity has also been predicted to occur in the $\nu=0$ quantum Hall state of graphene\cite{Takei2016}. As a matter of fact, in this state graphene is in an easy-plane canted antiferromagnetic phase\cite{Young2012}, which supports spin superfluid transport as experimentally observed recently \cite{Stepanov2018}.

Inspired by these results, the present work aims to better understand the spin transport properties of non-collinear antiferromagnetic insulators. In fact, non-collinear antiferromagnets host a wealth of intriguing properties. Non-collinear antiferromagnets displaying non-coplanar spin configuration exhibit anomalous Hall transport in the absence of spin-orbit coupling due to the emergence of a non-vanishing Berry curvature\cite{Shindou2001,Martin2008,Ndiaye2019}, but also anomalous magnonic transport \cite{Li2016g,Jian2018,Owerre2018} as well as exotic excitations in the quantum limit\cite{Shimizu2003,Yamashita2009,Shen2016,Clark2019}. In recent years, intense effort has been paid on non-collinear {\em coplanar} antiferromagnets with the prediction and observation of anomalous Hall effect\cite{Chen2014,Kubler2014,Nakatsuji2015,Nayak2016}, magneto-optical Kerr effect \cite{Higo2018} and magnetic spin Hall effect \cite{Zelezny2017b,Kimata2019} in Mn$_3$Ir and Mn$_3$Sn compounds. The classical non-collinear antiferromagnetic state has been studied in two-dimensional triangular and kagom\'e spin systems with Heisenberg exchange interactions and in-plane anisotropy \cite{Kurz2000,Heinze2002,Szunyogh2009,Liu2017b}. In the presence of easy-plane anisotropy, all magnetic moments have in-plane configuration and anomalous magnonic transport has been predicted \cite{Owerre2017e,Owerre2017b,Laurell2018,Mook2019,Kim2019d}. 

In this work, we investigate the spin transport across the magnetic phase diagram of a frustrated antiferromagnetic insulator and uncover a drastic modification of the transport regime across the various magnetic phases. To do so, we adopt a triangular lattice accounting for both nearest neighbor and next-nearest neighbor exchange interactions with easy-plane anisotropy (see Fig. \ref{Fig2}). By performing atomistic spin simulations on the two-terminal configuration depicted on Fig. \ref{Fig1}, we show that spin transport can be tuned from the diffusive to the spin superfluid regime depending on the level of the magnetic frustration. In particular, we show that non-collinear coplanar antiferromagnets exhibit spin superfluidity. When the frustration is strong enough to overcome the easy-plane anisotropy and cant the magnetic moment out of plane, the transport becomes diffusive. Finally, we also uncover the possibility for {\em proximate} spin superfluidity close to the ferromagnetic transition.

This paper is organized as follows. In Section \ref{SII} we introduce the Hamiltonian of the triangular spin system and its magnetic phase diagram. In Section \ref{SIII}, we compute the spin transport using atomistic spin simulation technique and discuss the emergence of the spin superfluid transport across the magnetic phase diagram. We show the analogy of magnetohydrodynamic equations of the non-collinear antiferromagnetic state with the supercurrent Josephson equations. Conclusion is given in Section \ref{SIV}.

\section{The Spin Model\label{SII}}

\begin{center}
\begin{figure}[h]
\includegraphics[width=12cm]{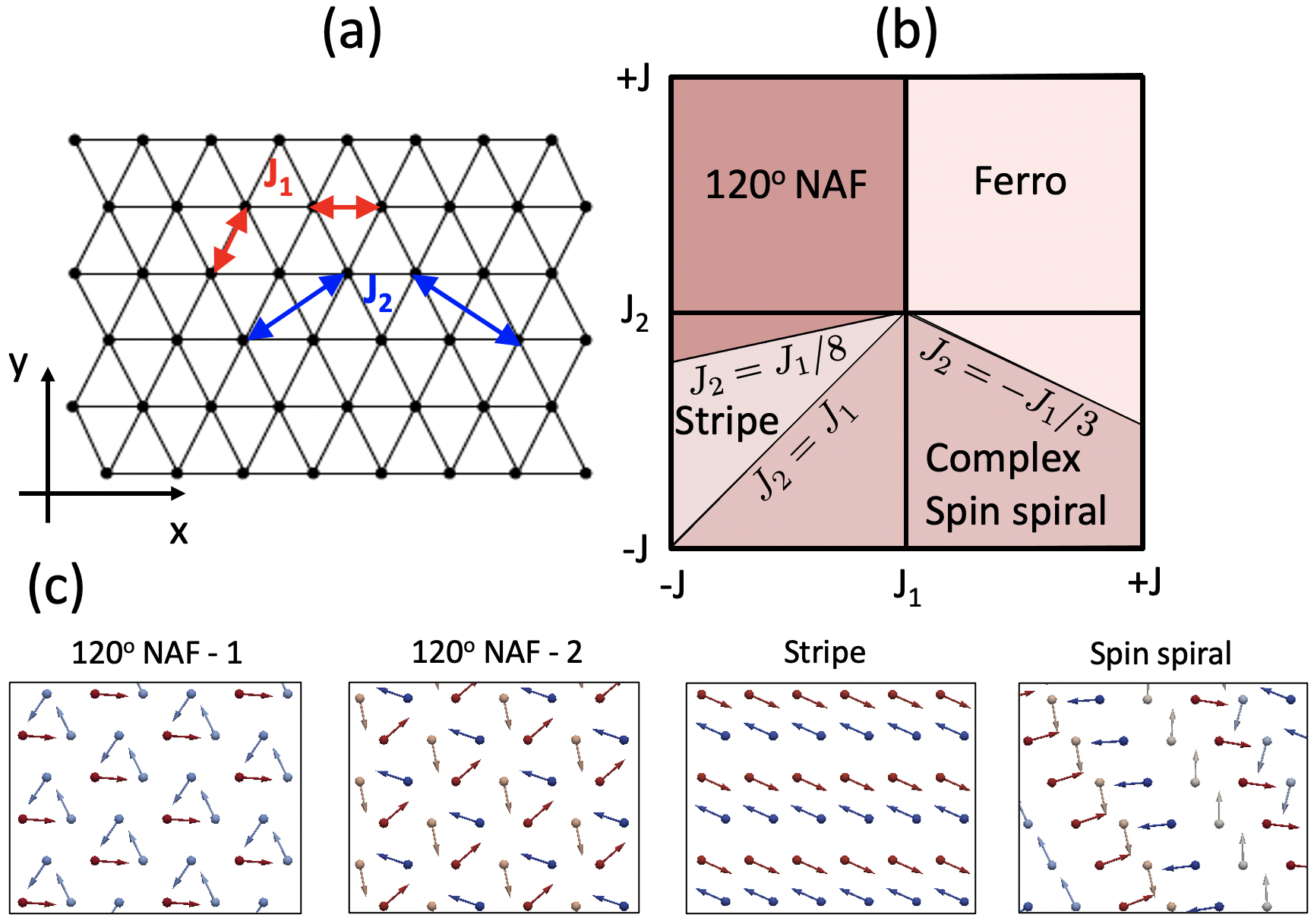}
\caption{(a) Schematics of the triangular lattice with nearest and next-nearest neighbor Heisenberg exchange. (b) Magnetic phase diagram obtained by atomistic spin simulations and (c) snapshots of the magnetic ground state of four representative configurations in the phase diagram. NAF-1 and NAF-2 correspond to $J_2>0$ and $J_2<0$, respectively.}
\label{Fig2}
\end{figure}
\end{center}

Our model system is a frustrated triangular system with easy-plane anisotropy represented on Fig. \ref{Fig2}(a). 
The Heisenberg spin Hamiltonian is given by
\begin{align} 
\mathcal{H}=-J_1 \sum_{\langle i,j\rangle}{\bf m}_i \cdot {\bf m}_j - J_2\sum_{\langle\langle i,j\rangle\rangle}{\bf m}_i \cdot {\bf m}_j+ K\sum_{i}({\bf m}_i \cdot \hat{z})^2,
\label{atomistic_hamiltonian}
\end{align} 
 where $\langle i,j\rangle$ and $\langle\langle i,j\rangle\rangle$ denote a sum over all nearest-neighbor and next-nearest-neighbor sites, respectively, ${\bf m}_i$ is the normalized unit spin vector at site $i$. The first two terms in Eq. \eqref{atomistic_hamiltonian} represent the nearest-neighbor and next-nearest-neighbor Heisenberg exchange interactions and the last term is the easy-plane anisotropy energy corresponding to hard-axis in the $z$-direction ($K>$0). The exchange parameters vary $-J \le$$J_1$,$J_2$$\le J$ in the units of $J$ within the $J_1-J_2$ phase space. The exchange interaction is ferromagnetic when $J_1 >$0 ($J_2>$0) and it is antiferromagnetic when $J_1 <$0 ($J_2<$0). \par

The magnetic frustration inherent to the triangular antiferromagnet inspired Anderson \cite{Anderson1973} to propose his resonating valence bond state and since then, this system has played a considerable role in the understanding of the geometrically frustrated classical and quantum systems \cite{Lacroix2011,Zhou2017}. In the classical limit, the competition between $J_1$ and $J_2$ stabilizes ferromagnetic, collinear and non-collinear antiferromagnetic and spin-spiral phases \cite{Kurz2000,Heinze2002}. To build the phase diagram, we use a home-made atomistic spin simulation code tracking the real-time dynamics of the triangular magnet with 90$\times$30 sites and periodic boundary conditions along both $x$- and $y$-directions. The $J_1$-$J_2$ phase diagram of the triangular lattice is shown in Fig. \ref{Fig2}(b) and typical snapshots of the magnetic configuration computed with our atomistic spin model are given in Fig. \ref{Fig2}(c). \par

When $J_1>0,J_2>-J_1/3$ (top right quadrant), the ferromagnetic state is stabilized. When $J_1<0,J_2>0$ (top left quadrant), a non-collinear coplanar antiferromagnetic phase is favored. In this phase, the three atomic sites from each sublattice occupy a ($\sqrt{3}\times\sqrt{3}$)R30$^\circ$ unit cell with a relative angle between nearest-neighbor spins of 120$^\circ$. Hence, this phase is called 120$^\circ$ N\'eel antiferromagnetic phase (NAF). Interestingly, the 120$^\circ$ NAF extends between $J_2=J_1/8$ and $J_2=0$. In fact, a quantum spin-liquid phase\cite{Zhou2017} has been predicted in this region (between the 120$^\circ$ NAF and stripe phases) by quantum Monte Carlo and density matrix renormalization group techniques \cite{Kaneko2014,Zhu2015,Hu2015b}. In the absence of easy-plane anisotropy, the quantum spin-liquid region locates between $0.06 \pm 0.1 \le J_2/J_1 \le 0.14 \pm 0.1$. Since our system assumes classical spins with easy-plane anisotropy, it does not display such a quantum spin liquid phase. When $J_2<{\rm min}\{J_1/8,-J_1/3\}$, the phase diagram is much richer. One can identify a large region, bounded between $J_2 = -J_1/3$ and $J_2=J_1$, where complex spin spiral states are obtained. Between the boundaries $J_2=J_1$ and $J_2=J_1/8$, a row-wise collinear antiferromagnetic phase, also called "stripe" phase, is stabilized.  The phase diagram depicted in Fig. \ref{Fig2} shows that the classical triangular Heisenberg magnet is a versatile platform to investigate the transmission of spin information in complex magnets as very different magnetic phases can be obtained out of frustration by simply tuning the relative strength of the magnetic exchange interactions.

\onecolumngrid
\twocolumngrid

\begin{figure}[htbp]
\begin{center} \includegraphics[width=09.5cm,height=07.5cm]{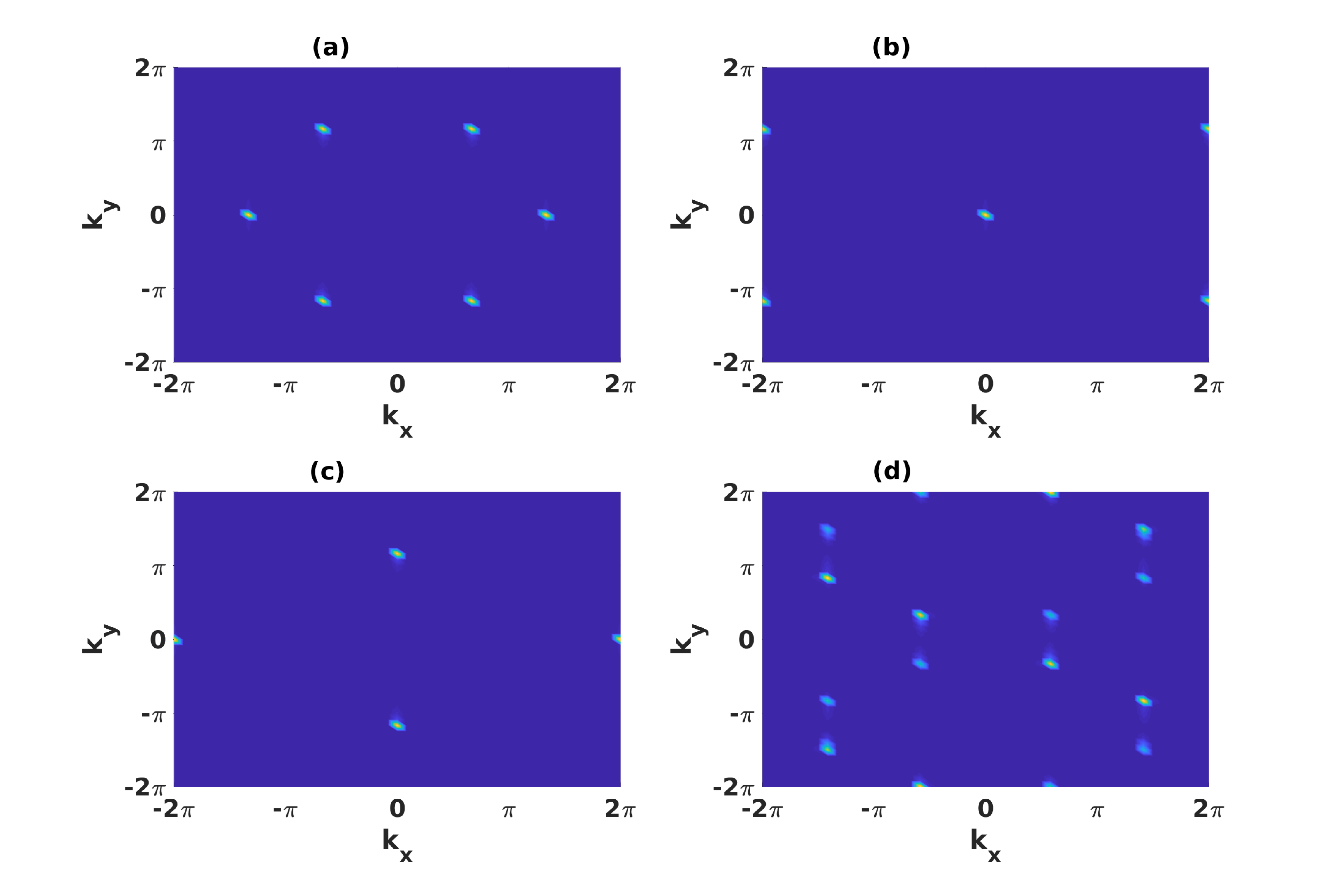}
\caption {Structure factor $S({\bf k})$ in momentum space for four magnetic phases of in the $J_1-J_2$ phase diagram. The peaks characterize long-range magnetic order.
(a) 120$^\circ$ NAF phase with ${\bf Q}$ = $\pm \frac{2\pi}{a}(\frac{1}{3},\frac{1}{\sqrt{3}})$,
$\pm \frac{4\pi}{a}(\frac{1}{3},0)$; (b) Ferromagnetic phase with ${\bf Q}=(0,0)$; (c) Stripe phase with ${\bf Q}$ = $\pm \frac{2\pi}{a}(0,\frac{1}{\sqrt{3}})$; (d) Example of a spin spiral phase showing complex but well-defined ${\bf Q}$  pattern.}
\label{figstrucfactor}
\end{center} \end{figure}

In order to confirm that the equilibrium state of a given phase matches the spin configuration of the ground state, we compute the spin structure factor of the state by solving the Landau-Lifshitz-Gilbert equation [see Eq. \eqref{llgeqnm}]. Using the discrete Fourier transform of the spin-spin correlation $\langle{\bf m}_i \cdot {\bf m}_j \rangle$ in the equilibrium state, the spin structure factor can be obtained by the following equation,
\begin{eqnarray}
S({\bf k})= \frac{1}{N}\sum_{ij} e^{i{\bf k}\cdot ({\bf r}_i-{\bf r}_j)} \langle{\bf m}_i \cdot {\bf m}_j \rangle,
\end{eqnarray}
 where ${\bf r}_i$ ,${\bf r}_j$ are the positions of spins ${\bf m}_i$ and ${\bf m}_j$ with total number of atoms $N$. Here we take $N = 90\times30$ sites for the spin structure factor calculations. In Fig. \ref{figstrucfactor}, we plot the spin structure factor in the ($k_x$, $k_y$) plane with the Bragg peaks for each of these phases. The peaks corresponding to $S({\bf k})$ in the first Brillouin zone indicates the wave vector ${\bf Q}$ of a given phase. The wave vector corresponding to the 120$^\circ$ NAF phase is shown in Fig. \ref{figstrucfactor}(a), described by ${\bf Q}$ = $\pm \frac{2\pi}{3a}(\frac{1}{3},\frac{1}{\sqrt{3}})$ and $\pm \frac{4\pi}{a}(\frac{1}{3},0)$. The wave vector representing the $\Gamma$ point in the Fig. \ref{figstrucfactor}(b), ${\bf Q}$ = (0,0) indicates the ferromagnetic phase. The wave vector corresponding to the stripe phase is shown in Fig. \ref{figstrucfactor}(c), represented by ${\bf Q}$ = $\pm \frac{2\pi}{a}(0,\frac{1}{\sqrt{3}})$. We observe a change in the structure factor peaks from 120$^\circ$ NAF to stripe phase at $J_2=J_1/8$ while coming from 120$^\circ$ NAF to stripe phase. The spin structure factor for a representative spin spiral is shown in Fig. \ref{figstrucfactor}(d), displaying clear peaks that correspond to well defined wave vectors. The structure factor calculation of the ferromagnetic, 120$^\circ$ N\'eel and stripe phases are in agreement with the semi classical \cite{Jolicoeur1990} and quantum approaches \cite{Bishop2015,Maksimov2019}.

\section{Spin transmission in the triangular magnet\label{SIII}}
\subsection{Atomistic spin simulations}
To simulate the spin transport properties, we consider atomic spins arranged on a triangular lattice with 90$\times$30 sites, as described in the previous section, and perform atomistic spin simulations \cite{Evans2014}. To make our result physically sounding, we adopt a lattice constant, $a$ = 3 \AA~. We set the exchange parameters as -$J\le J_1,J_2\le J$ and impose a fixed easy-plane anisotropy $K$=0.01 $J$, with $J=1$ meV. We adopt the device structure depicted on Fig. \ref{Fig1}, namely the magnetic layer is extended along the $x$-direction, embedded between two metallic electrodes placed at $x=0$ and $x=L$. In addition, periodic boundary conditions are applied along the $y$-direction in order to minimize finite size effects. As explained above, the charge current flowing in the left electrode injects a spin current into the first column of spins on the left side of the magnetic layer via the spin Hall effect. On the left side, the equation of motion of a given spin $i$ is given by
\begin{align}
\partial_t{\bf m}_i=&-\gamma {\bf m}_i \times {\bf H}_i^{\rm eff} +(\alpha_0 + \alpha^\prime) {\bf m}_i \times \partial_t{\bf m}_i \nonumber \\
&-\alpha^\prime {\bf m}_i \times ({\bf m}_i \times {\boldsymbol \mu}_L/\hbar).
\label{llgeqnl}
\end{align}
Here, $\gamma$ is the absolute value of the gyromagnetic ratio, the effective field of $i^{th}$ spin is represented by ${\bf H}_i^{\rm eff}=-\frac{1}{\mu_s}{\partial_{{\bf m}_i}}\mathcal{H}$. The energy dissipation is given by the intrinsic Gilbert damping, $\alpha_0$, and the interfacial damping\cite{Tserkovnyak2002}, $\alpha^\prime$. The last term accounts for the spin transfer torque exerted by the input spin accumulation ${\bm \mu}_L$. Here $\alpha^\prime$ = $\frac{g^{\uparrow\downarrow}\hbar^2\gamma}{2e^2M_sa}$ \cite{Skarsvag2015} is the interfacial damping associated with interfacial spin pumping, where $g^{\uparrow\downarrow}$ is the spin mixing conductance. On the opposite side of the magnetic layer, the precessing spins pump a spin current into the right electrode, which is eventually converted into an electric signal via inverse spin Hall effect. There, the Landau-Lifshitz-Gilbert equation of the interfacial spins reads 
\begin{align}
\partial_t{\bf m}_i=&-\gamma {\bf m}_i \times {\bf H}_i^{\rm eff} +(\alpha_0 + \alpha^\prime) {\bf m}_i \times \partial_t{\bf m}_i \nonumber \\
&-\alpha^\prime {\bf m}_i \times ({\bf m}_i \times {\boldsymbol \mu}_R/\hbar),
\label{llgeqnr}
\end{align}
where $\alpha'$ accounts for the interfacial spin pumping in the right electrode. In the rest of the magnetic layers, the equation of motion is given by 

\begin{eqnarray}
\partial_t{\bf m}_i=-\gamma {\bf m}_i \times {\bf H}_i^{\rm eff} +\alpha_0 {\bf m}_i \times \partial_t{\bf m}_i.
\label{llgeqnm}
\end{eqnarray}

The output is given by the spin accumulation pumped by the precessing magnetic moments along the right boundary, ${\bm\mu}_R=-\hbar \sum_{i\in R} ({\bf m}_i \times \partial_t {\bf m}_i)$, as derived from interfacial spin pumping theory \cite{Skarsvag2015,Qaiumzadeh2017,Tserkovnyak2005}. Only the perpendicular component of the interfacial spin accumulation $\mu_R^z$ reaches a steady value whereas the in-plane components, $\mu_R^x$ and $\mu_R^y$, are much smaller ($\sim 10^{-3}$) and oscillate in time. In the following, we focus on the perpendicular component $\mu_R^z$.

The spin mixing conductance is set to $g^{\uparrow\downarrow}\approx(e^2/h)10^{18}$ m$^{-2}$, comparable to the one measured in Pt$|$YIG$|$Pt structure \cite{Qiu2013,Baltz2018}, and the saturation magnetization is taken similar to that of YIG, $4\pi M_s$ = 1750 G \cite{Serga2010}. With these parameters, $\alpha^\prime$ $\approx$ 0.1, much larger than the intrinsic Gilbert damping expected in magnetic insulators $\alpha_0\approx$ 10$^{-3}$. We emphasize that the current flowing in the injector is continuous and that our calculations are performed at zero temperature. Therefore, the torque exerted at the left boundary does not excite thermal magnons nor does it induce ferromagnetic resonance. This is the ideal setup to observe spin superfluidity, if any.\par

 \begin{figure}[htbp]
\begin{center} \includegraphics[width=9cm]{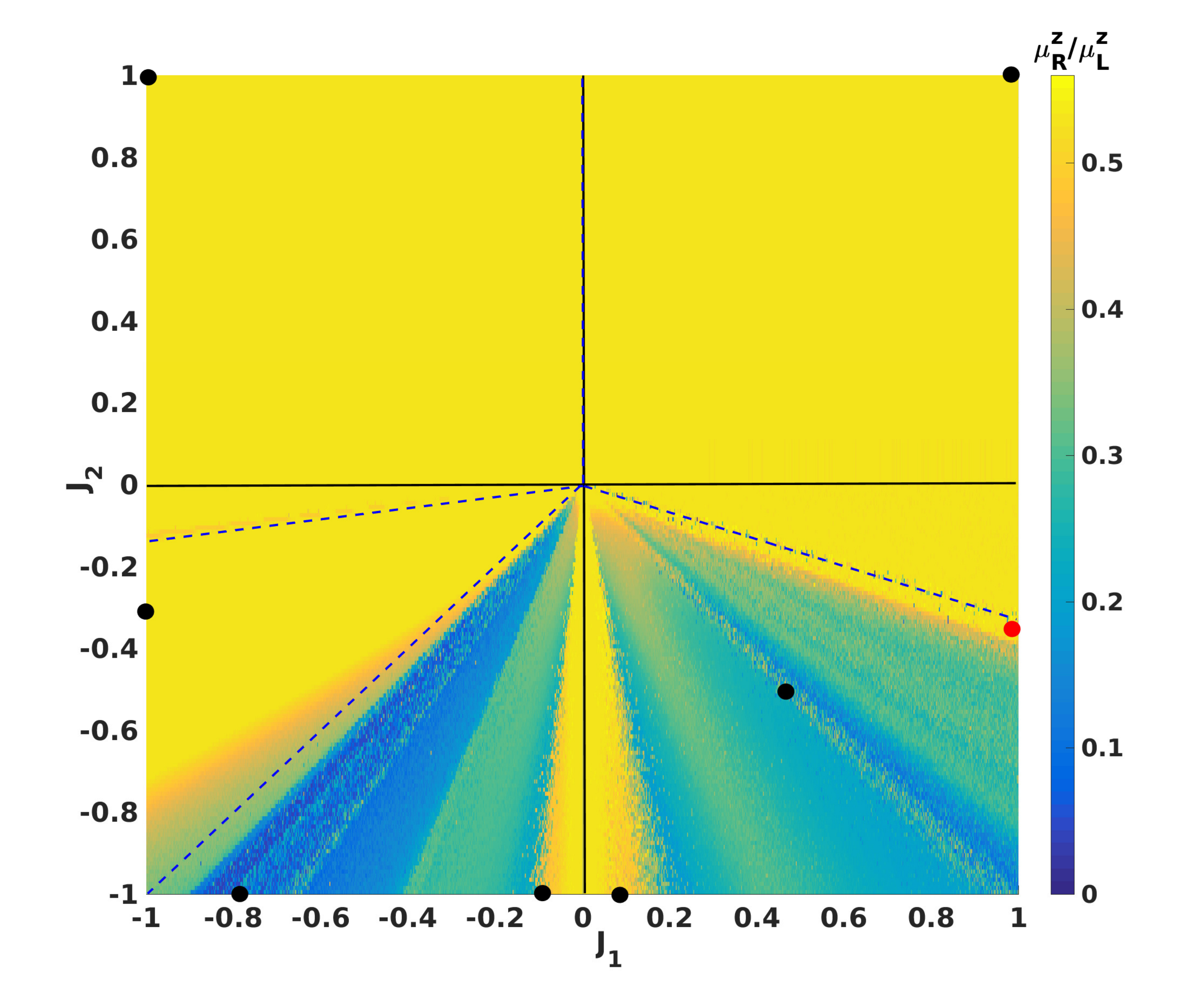}
\caption{ The spin accumulation ratio ($\mu_R^z$/$\mu_L^z$) of all four phases of 90$\times$30 sites in $J_1-J_2$ phase diagram. The black and red dots indicate different ($J_1,J_2$) configurations for which length-dependent spin transport properties have been calculated. $J_1$ and $J_2$ are in units of $J=1$ meV. All phase boundaries are indicated by blue dotted lines.}
\label{spinaccumulation_diagram}
\end{center} \end{figure}

\subsection{Spin accumulation phase diagram}
To understand how spin information is transmitted in this device, we have systematically computed the system's output $\mu_R^z/\mu_L^z$ across the magnetic phase diagram. For computational purposes, the phase diagram is divided into 1001 $J_1$ and 201 $J_2$ points in the range -1 to 1 corresponding to 1001$\times$201 grid points. First, for each set of $J_1$ and $J_2$ the magnetic ground state configuration is first computed in the absence of spin injection ($\mu_L^z=0$).  To reach the ground state, the time evolution of Eq. \eqref{llgeqnm} is solved while assuming random spin configuration as an initial state. The ground state of each phase is confirmed by accomplishing equilibrium spin configuration and energy convergence as discussed in Section \ref{SII}. In the presence of non-zero $\mu_L^z$, the spin accumulation exerts a torque on the interfacial spins and excites collective magnetization dynamics. Then the spins are allowed to evolve until steady state is attained. The resulting state has a canted spin configuration with a non-zero $m_z$. We consider the system has reached a steady state when $m_z$ becomes time-independent. In this state, the spin moments exhibit persistent in-plane rotation due to the continuous spin injection occurring at the left boundary.
In order to ensure saturation of $m_z$ and $\mu_R^z$, simulations are carried out up to 50 ns of numerical integration time for all phases. We take $\mu_L^z=10^{-3}$ meV to carry out the spin transport calculations, ensuring that our simulations are conducted in the linear response regime ($\mu_R^z\propto \mu_L^z$). The ratio $\mu_R^z/\mu_L^z$ is plotted as a function of $J_1$ and $J_2$ in Fig. \ref{spinaccumulation_diagram} for all phases for 90$\times$30 sites. The black lines mark the phase boundaries determined in Fig. \ref{Fig2}.

\begin{figure}[htbp]
\vspace{1.0cm}
\begin{center} \includegraphics[width=8.5cm,height=6.0cm]{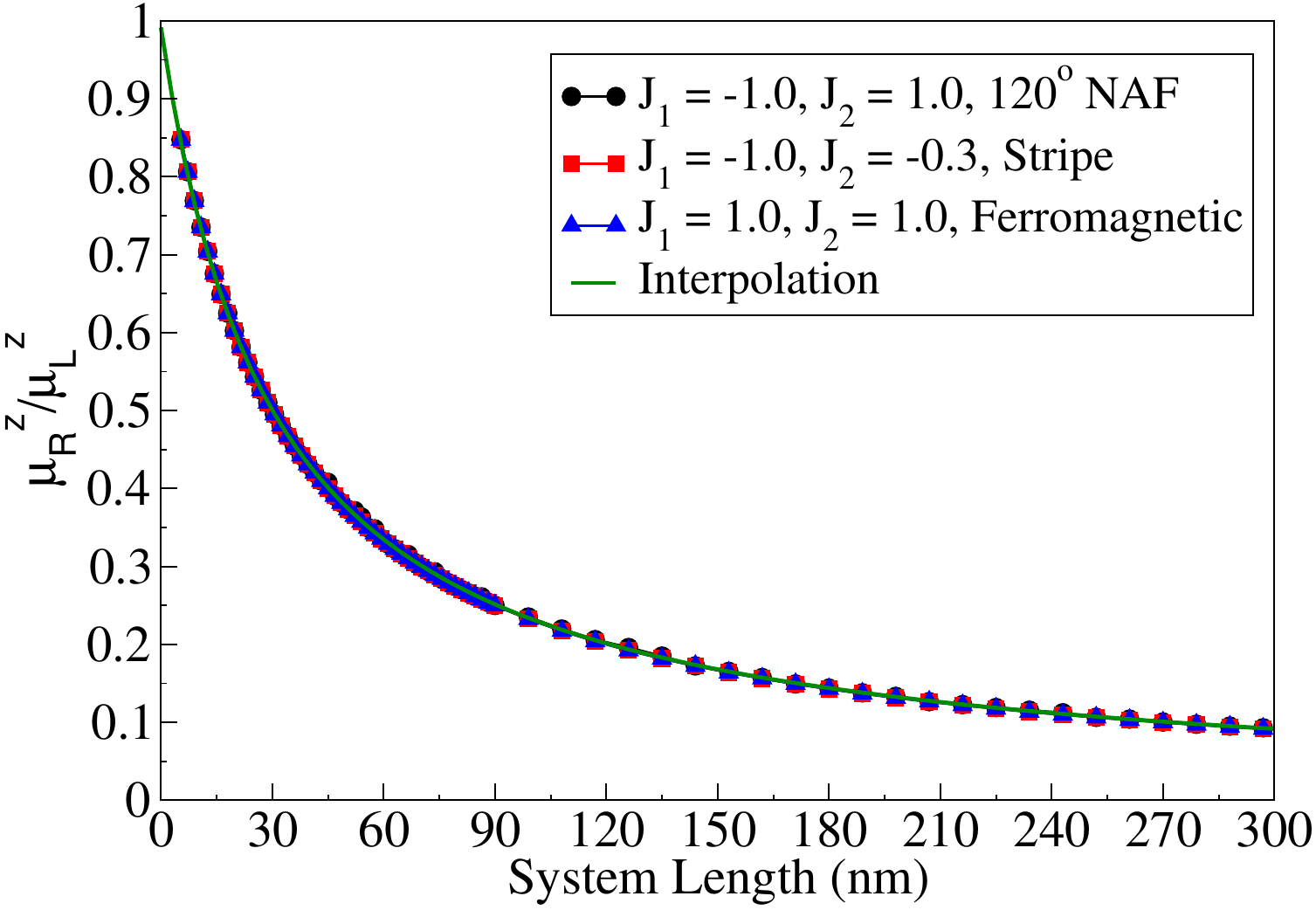}
\caption{The spin accumulation ratio ($\mu_R^z$/$\mu_L^z$) as a function of triangular system length L. The exchange interaction strengths are given in the inset. The interpolation curve is based on the 120$^\circ$ NAF and stripe phase curves. $J_1$ and $J_2$ are in units of $J=1$ meV.}
\label{lengthdependence1}
\end{center} \end{figure}

We first consider the three phases that display maximum output, $\mu_R^z/\mu_L^z=0.526$ (yellow region): 120$^\circ$ NAF, stripe and ferromagnetic phases. The ferromagnetic phase ($J_1>0,J_2>-J_1/3$) displays the spin superfluid behavior expected in the absence of long-range dipolar interaction \cite{Sonin2010,Takei2014,Skarsvag2015}. In the case of the stripe phase ($J_1<J_2<J_1/8<0$), the magnetic phase is represented by a single N\'eel vector that precesses about the $z$-axis and therefore experiences spin superfluidity, consistently with previous prediction \cite{Takei2014b}. More remarkably, a similar behavior is observed in the 120$^\circ$ NAF phase ($J_1<0$, $J_1/8<J_2$), where the magnetic configuration is described by two N\'eel vectors ${\bf l}_1$ and ${\bf l}_2$ that are constrained in-plane and precess about the $z$-axis. Clearly, the output spin accumulation is independent of the value of $J_1$ and $J_2$ exchange interactions, which is consistent with the spin superfluidity expected in planar antiferromagnetic insulators \cite{Takei2014b}. Notice that at the boundary between the 120$^\circ$ NAF and stripe phases, we observe small fluctuations of the output (not shown), attributed to the weak stability of this region, as discussed above.\par

The nature of long-range spin transport reveals the possibility of spin superfluid transport in magnetic insulators. We calculate the spin accumulation ratio for different system sizes from each magnetic phase and report it as a function of the system length in Fig. \ref{lengthdependence1}. For these long-range spin transport calculations, we have taken system size as $N_x\times$30 with $N_x$ increasing from 18 sites to 990 sites. The corresponding system length varies from 5.4 nm to 297 nm. From this figure, the spin accumulation ratio $\mu_R^z$/$\mu_L^z$ for the 120$^\circ$ NAF, stripe and ferromagnetic phases all collapse on the same algebraic decay (solid green line), independently on the value of the exchange parameters. This result demonstrates long-range spin superfluid spin transport in the 120$^\circ$ NAF phase. The corresponding fitting equation is $\frac{\mu_R^z}{\mu_L^z}$ = $\frac{1}{1+0.033L}$.

\begin{figure}[htbp]
\vspace{1.0cm}
\begin{center} \includegraphics[width=8.5cm,height=6.0cm]{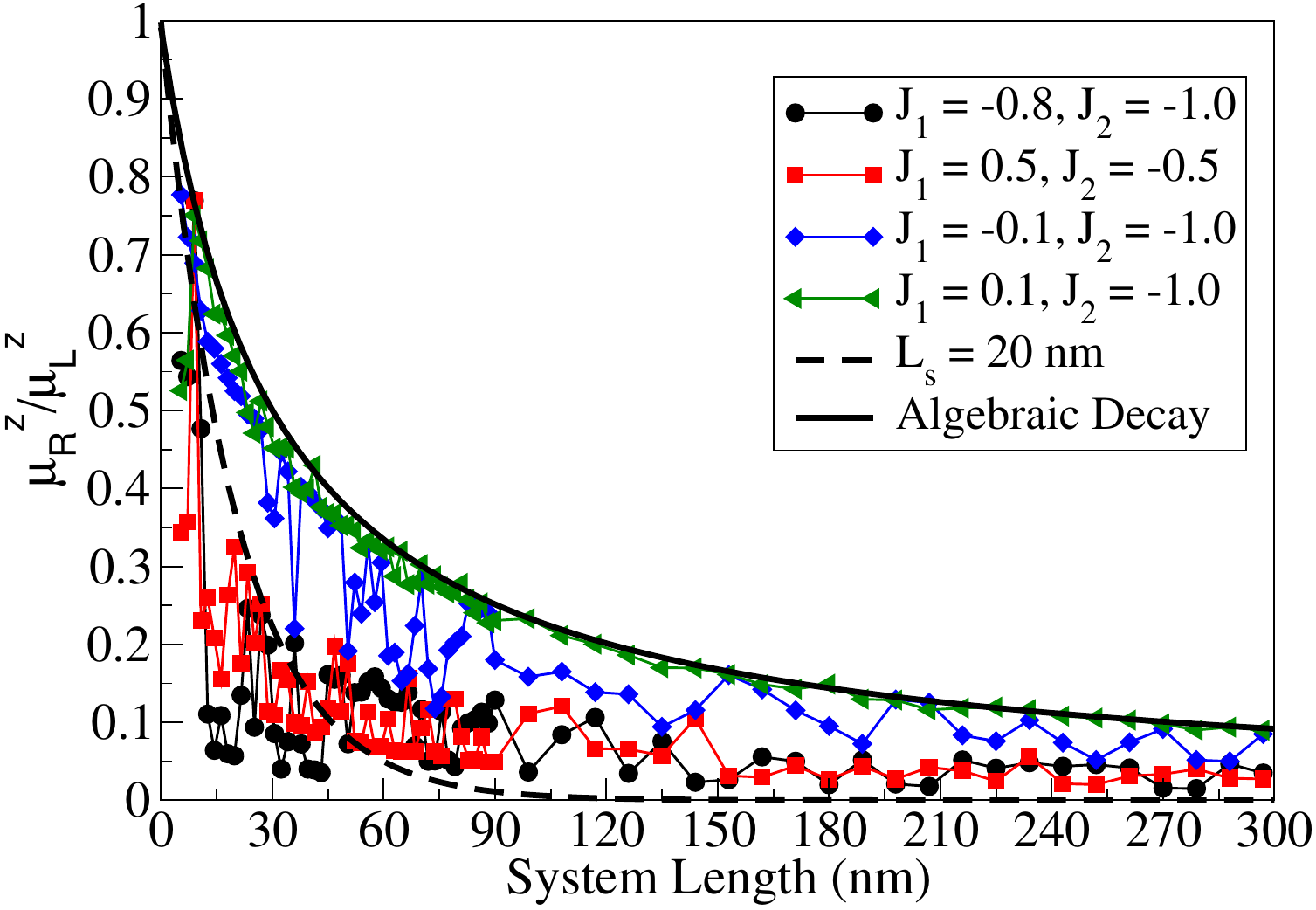}
\caption{The spin accumulation ratio ($\mu_R^z$/$\mu_L^z$) as a function of triangular system length L. 
The exchange interaction strengths are given in the inset. The interpolation curve is based on the 120$^\circ$ NAF and stripe phase curves. $J_1$ and $J_2$ are in units of $J=1$ meV.}
\label{lengthdependence2}
\end{center} \end{figure}

Let us now turn our attention on the last region of the diagram, occupied by complex, possibly inhomogeneous spin spirals ($J_2<{\rm min}\{J_1,-J_1/3\}$). In this region, the competition between nearest-neighbor and next-nearest neighbor exchange interactions promotes a range of highly frustrated non-collinear magnetic configurations with spin moments pointing out of plane. As a result, the output displays a non-monotonic behavior in the phase space. In the limit $|J_2|\gg |J_1|$, the out-of-plane canting of the spin moments is weak and therefore the spins dynamics is close to that of easy-plane magnets, yielding a maximum output $\mu_R^z$/$\mu_L^z\approx0.5$. However, away from this line, the out-of-plane canting increases drastically, which violates the in-plane nature of spin superfluidity. Though the steady state is achieved, the resultant output is severely reduced.
Similarly, at the boundary between stripe ($J_2=J_1/8$) and ferromagnetic phases ($J_2=-J_1/3$), the magnetic ground is only marginally stable and the dynamics deviates from the spin superfluid behavior encountered away from the boundaries.\par

Figure \ref{lengthdependence2} displays the dependence of the spin accumulation output as a function of the length of the magnet, for four characteristic sets of parameters as indicated by the black dots in Fig. \ref{spinaccumulation_diagram}: two points are taken close to $J_1=\pm0.1 J$ (green and blue), where the output is close to maximum, and two points are taken further away in the spin spiral region (black and red). Close to $J_1=0$ axis (green and blue), we observe that the decay is fairly close to algebraic (solid line), revealing that the spin transport regime is a {\em proximate} spin superfluid. Away from this axis (black and red), the decay is dramatically enhanced and is qualitatively described by an exponential decay, characteristic of {\em diffusive transport}. We emphasize that we ensured that the data reported on Fig. \ref{spinaccumulation_diagram} is well converged numerically, so that the large fluctuations of the blue, red and black curves are not due to convergence issues but rather to finite size effects. As a guide for the eye, the dashed line represents an exponential decay ($~\sim e^{-L/L_s}$) with a spin diffusion length $L_s$=20 nm.\par

\begin{figure}[htbp]
\vspace{0.5cm}
\begin{center} \includegraphics[width=8cm]{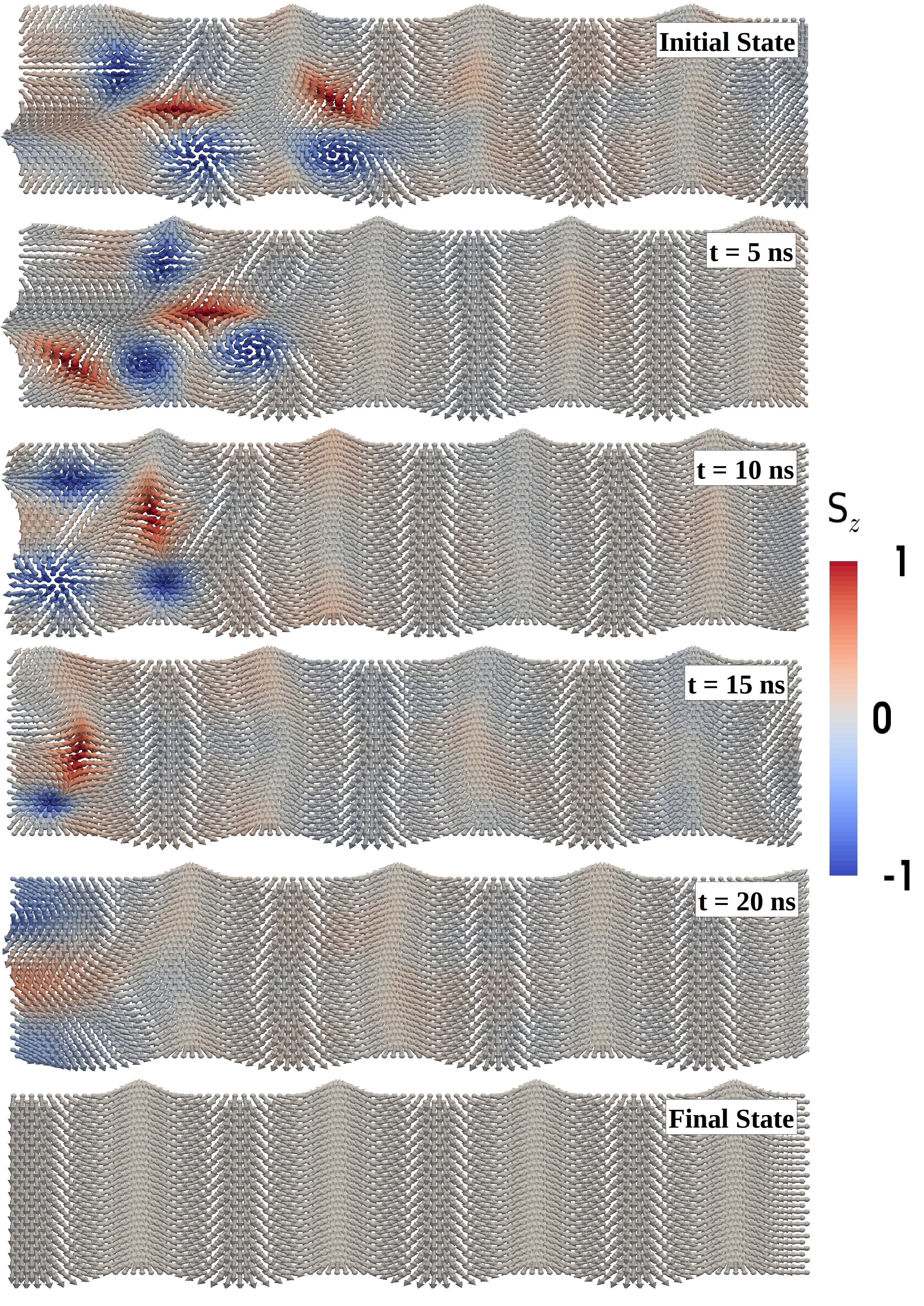}
\caption{Example of a complex, inhomogeneous spin spiral ground state, located at $J_1 = J$ and $J_2 = -0.34J$, close to the boundary with the ferromagnetic state. The high frustration arising from the competition between nearest-neighbor and next-nearest neighbor exchange interactions results in a highly degenerate ground state. In spite of its high inhomogeneity, this complex magnetic texture is able to transmit spin information almost perfectly due to its proximity with the planar ferromagnetic state.}
\label{spinspiral}
\end{center} \end{figure}

Finally, an interesting feature of the complex, inhomogeneous spin spiral phase is worth mentioning. We find that in proximity to the planar ferromagnetic phase, the output is close to maximum in spite of the complex magnetic texture (see red dot in Fig. \ref{spinaccumulation_diagram}). As an illustration, we consider the magnetic state taken at $J_1 = J$ and $J_2 = -0.34J$ and indicated by the red spot on Fig. \ref{spinaccumulation_diagram}. The ground state is a quasi-one dimensional planar spin spiral extended along the length of the magnetic sample. Due to the strong magnetic frustration, we find that other nearly-degenerate complex spin spiral states can be stabilized, depending on the initial conditions. For instance, one of these states consists in two magnetic vortices sitting on the quasi-one dimensional spin spiral, as reported on Fig. \ref{spinspiral}. Since the energy difference between these excited states and the true ground state is very small, they are likely to emerge naturally in real frustrated magnets. 

We now consider the evolution of this particular state under spin injection. Figure \ref{spinspiral} presents snapshots of the magnetic texture between the initial state ($t=0$ ns) and the final state ($t=50$ ns). Upon turning on the spin injection, the magnetic texture is progressively modified, the vortices are eventually expelled and a planar spin spiral akin to the ground state finally stabilizes. As a result, in spite of the complexity of the initial state and of the presence of out-of-plane spin orientation, the final spin transmission remains remarkably efficient. These simulations demonstrate that dc pumping of spin currents in non-collinear magnetic insulators can occur via either diffusive or spin superfluid regime and that magnets displaying weakly out-of-plane spin moments can display proximate spin superfluid signature.

\subsection{Spin superfluidity in non-collinear antiferromagnet\label{ssp120}}

Whereas the atomistic model provides an accurate description of the spin dynamics, it lacks transparency. To complete this study and better understand the conditions under which spin superfluidity emerges in the 120$^\circ$ NAF phase, an analytical theory based on the continuous micromagnetic approximation is more appropriate, as demonstrated in the case of spin superfluid transport in easy-plane ferromagnetic and antiferromagnetic systems \cite{Takei2014,Qaiumzadeh2017}. In this approximation, the atomistic Heisenberg Hamiltonian Eq. \eqref{atomistic_hamiltonian} translates to the free energy
$F = \int(f_{ex} + f_{ani})d^3r$. Here $f_{ex}$ and $f_{ani}$ are the exchange and
magnetic anisotropy energies of each sublattice \cite{Kosevich1990} and they are expressed as
\begin{flalign}
f_{ex} =& J({\bf m}_A \cdot {\bf m}_B + {\bf m}_B \cdot {\bf m}_C + {\bf m}_C \cdot {\bf m}_A) \nonumber \\
& +  \frac{A_1}{3}[({\bf \nabla m}_A)^2 + ({\bf \nabla m}_B)^2 + ({\bf \nabla m}_C)^2] \nonumber \\
& +  \frac{2A_2}{3}[({\bf \nabla m}_A\cdot{\bf \nabla m}_B) + ({\bf \nabla m}_B\cdot{\bf \nabla m}_C) \nonumber \\
& + ({\bf \nabla m}_C\cdot{\bf \nabla m}_A)], \\
f_{ani} =& \frac{K}{3} [({\bf m}_A \cdot {\hat z})^2 + ({\bf m}_B \cdot {\hat z})^2  +  ({\bf m}_C \cdot {\hat z})^2],
\end{flalign}
where ${\bf m}_A$, ${\bf m}_B$ and ${\bf m}_C$ are the spin moments in each sublattice. 
The exchange interaction strength is $J$, inhomogeneous stiffness constants are $A_1$ and $A_2$ respectively 
and $K$ is the strength of hard-axis anisotropy. The equation of motion of the each sublattice $i$ is then given by Eq. \eqref{llgeqnm}, with $\partial_t{\bf m}_i=(\gamma/M_sV) {\bf m}_i\times\delta F/\delta {\bf m}_i+\alpha {\bf m}_i \times \partial_t{\bf m}_i$, $M_s$ being the saturation magnetization and $V$ the volume of the magnet.

To describe the soft modes of the three-sublattice antiferromagnet, we introduce 
the two N\'eel vectors ${\bf l}_1 = \frac{1}{3{\sqrt 2}}[{\bf m}_A + {\bf m}_B - 2{\bf m}_C]$ and
${\bf l}_2 = \frac{1}{{\sqrt 6}}[-{\bf m}_A + {\bf m}_B]$, and
magnetization ${\bf m} = \frac{1}{3}[{\bf m}_A + {\bf m}_B + {\bf m}_C]$ \cite{gomonay2015}.
To conserve sublattice spin moments, the N\'eel vectors and the magnetization satisfy
${\bf l}_1^2+{\bf l}_2^2+{\bf m}^2=1$, ${\bf l}_1 \perp {\bf l}_2$ and
${\lvert {\bf l}_1 \rvert} = {\lvert {\bf l}_2 \rvert}$. In the strong exchange limit, 
the magnetization is considered to be small, ${\lvert {\bf m} \rvert} \ll 1$. 
In terms of the N\'eel vectors and magnetization, the free energy reads 
\begin{align} 
f =& \frac{9J}{2}{\bf m}^2 + A [({\bf \nabla l}_1)^2 + ({\bf \nabla l}_2)^2] 
+  K [({\bf l}_{1} \cdot {\hat z})^2+ ({\bf l}_{2} \cdot {\hat z})^2], \nonumber \\
\end{align}
where $A = A_1 - A_2$. By introducing ${\bf n} = {\bf l}_1 \times {\bf l}_2$, we can show that the magnetization 
is a slave variable of the N\'eel vectors and it can be written as  
\begin{equation}
{\bf m} = \frac{1}{\omega_J}[\partial_t{\bf l}_1 \times {\bf l}_1 + \partial_t{\bf l}_2 \times {\bf l}_2 + 
\partial_t{\bf n} \times {\bf n}],
\label{slave_magnetization}
\end{equation}
where $\omega_J=\frac{M_s}{18 \gamma J}$. Hence, in the strong exchange limit, the study of three-sublattice dynamics is now reduced to the dynamics of the two coupled N\'eel vectors. To obtain the equation of motion of the N\'eel vectors in the presence of spin accumulation, we express the Landau-Lifshitz-Gilbert equation in terms of magnetization as 
\begin{flalign}
\partial_t{\bf m}~=~&-\gamma [{\bf l}_1 \times {\bf H}_{{\bf l}_1}^{eff} ~+~ {\bf l}_2\times{\bf H}_{{\bf l}_2}^{eff} 
~+~ {\bf m}\times{\bf H}_{{\bf m}}^{eff}] \nonumber \\
& ~+~\alpha[{\bf l}_1 \times \partial_t{\bf l}_1 ~+~ {\bf l}_2 \times \partial_t{\bf l}_2] 
 - 3\alpha^\prime {\bf l}_1 \times ( {\bf l}_1 \times {\boldsymbol \mu}_L) \nonumber \\
&  - 3\alpha^\prime {\bf l}_2 \times ( {\bf l}_2 \times {\boldsymbol \mu}_L),
\label{llg_magnetization}
\end{flalign}
where ${\bf H}_{{\bf l}_1}^{eff}$, ${\bf H}_{{\bf l}_2}^{eff}$ and ${\bf H}_{{\bf m}}^{eff}$ are the effective fields of vectors ${\bf l}_1$, ${\bf l}_2$ and ${\bf m}$.
By substituting Eq. \eqref{slave_magnetization} in the above equation and simplifying, we get
\begin{flalign}
&{\bf l}_1 \times [\partial_t^2 {\bf l}_1 - \omega_J\omega_A \lambda^2 {\nabla}^2{\bf l}_1 
 + \omega_J\omega_K l_1^z \hat{z} + \omega_J \alpha \partial_t{\bf l}_1 \nonumber \\
&- \alpha^\prime {\bf l}_1 \times {\boldsymbol \mu}_L] 
 + {\bf l}_2 \times [\partial_t^2{\bf l}_2 - \omega_J\omega_A \lambda^2 {\nabla}^2{\bf l}_2 
 + \omega_J\omega_K l_2^z \hat{z} \nonumber \\ 
&+ \omega_J \alpha \partial_t{\bf l}_2 - \alpha^\prime {\bf l}_2 \times {\boldsymbol \mu}_L]
 + {\bf n} \times \partial_t^2{\bf n} = 0,
\label{eqn_neel_dynamics}
\end{flalign}
where $\omega_A \lambda^2=\frac{2\gamma A}{M_s}$ with domain wall length $\lambda$ and $\omega_K=\frac{2\gamma K}{M_s}$.
By considering the invariance of triangular lattice in ($y,z$)-plane, Eq. \eqref{eqn_neel_dynamics} can be solved for one-dimensional solutions along $x$-axis. 
We take ${\boldsymbol \mu}_L=\mu_L^z \hat{z}$ and using spherical coordinates, the N\'eel vectors can be parameterized as 
${\bf l}_1 = \frac{1}{\sqrt{2}}(l_1\cos\phi, l_1\sin\phi, l_1^z)$
and ${\bf l}_2 = \frac{1}{\sqrt{2}}(-l_2\sin\phi, l_2\cos\phi, l_2^z)$, where $\phi$ is azimuthal angle and $l_1=\sqrt{1-l{_1^z}^2}$ 
and $l_2=\sqrt{1-l{_2^z}^2}~$ along with $l{_1^z}^2,l{_2^z}^2 \ll 1$ \cite{Yamane2019}. To obtain the stable state solution, we solve Eq. \eqref{eqn_neel_dynamics} 
in the static limit. Considering only linear order in $l_1^z$, $l_2^z$ and gradients of $\phi$,
we get
\begin{align}
\omega_A \lambda^2 \partial_x^2 \phi &~=~ \frac{\alpha^\prime}{\hbar} \mu_L^z.
\label{static_neel1} 
\end{align}

From Eq. \eqref{static_neel1}, only one spatially varying solution exists. The solution is an in-plane homogeneous spiral state 
and this state is stable for non-zero $\mu_L^z$. This stable solution is in agreement with the steady state of antiferromagnetic systems in the absence of uniaxial anisotropy and magnetic field \cite{Qaiumzadeh2017}. From Eq. \eqref{static_neel1}, it is clear that a non-zero $\mu_L^z$ can excite magnetization dynamics.

Analogous to conserved quantities in the conventional superfluidity of $^4$He \cite{Kapitza1938,Volovik2003,Bunkov2013}
such as mass current, the spin superfluidity in easy-plane magnetic insulators refers to a pair of
canonical variables ($\phi$, $m_z$) with phase of condensation $\phi$ and transverse spin component $m_z$ \cite{Sonin2010}. 
Taking only the linear orders of $l_1^z$ and $l_2^z$ and derivatives of $\phi$, the dynamics of the canonical pair is represented as 
\begin{subequations}
\begin{align}
\partial_t{\phi} &= -\omega_J m_z,
\label{steady_eqn1}\\ 
\partial_t{m}_z &= -2\omega_A\lambda^2 \partial_x^2 \phi~+~2\alpha \dot{\phi}.
\label{steady_eqn2}
\end{align}
\end{subequations}
In the absence of Gilbert damping, the above coupled equations are a magnetic analog to the Josephson relations \cite{Sonin2010,Halperin1969}. This analogy suggests the possibility of spin superfluid transport through non-collinear antiferromagnetic system, as proven numerically in the previous section. In the steady state, $m_z$ is constant and $\phi(x,t)=\phi(x)+\Omega t$ are the solutions of Eqs. \eqref{steady_eqn1} and \eqref{steady_eqn2} \cite{Takei2014b}. Here $\Omega$ is the precession frequency and it depends on $\mu_L^z$ \cite{Qaiumzadeh2017}. 
From Eq. \eqref{steady_eqn2}, the spin current density is determined as 
$J_s(x)~=~2M_s\omega_A\lambda^2 \partial_x \phi(x)$.

\section{Conclusion\label{SIV}}
Using a frustrated triangular antiferromagnet as a simulation platform, we have studied the crossover from spin superfluid to diffusive spin transport. We have demonstrated that magnetic systems characterized by a strong easy-plane anisotropy support spin superfluid transport irrespective of their specific magnetic configuration (ferromagnet, collinear and non-collinear antiferromagnet), which results in a high device output independent of the values of nearest of next-nearest exchange parameters. The spin superfluidity is robust as long as the magnet remains away from magnetic phase transition boundaries, consistently with previous theoretical predictions\cite{Sonin2010,Konig2001,Takei2014}. \par

When approaching these boundaries, the transport slowly transits towards a diffusive regime that is fully reached in highly frustrated phases. At the boundary with the antiferromagnetic phase, the output starts deviating from the ideal spin superfluid transport even before reaching the magnetic phase transition as the stability of the coplanar ground state weakens. Nonetheless, our simulations also demonstrate a certain tolerance of the spin superfluid transport for deviations from coplanarity. When the frustration that allows for the spin moments to cant out of plane remains reasonable (e.g., close to the $J_1=0$ or to the $J_2=-J_1/3$ axes), the spin transport can re-enter the spin superfluid regime in spite of the complex non-collinear non-coplanar magnetic texture in this region.

\bibliography{Biblio2020}

\end{document}